\documentclass[aps,prl,reprint,groupedaddress]{revtex4-1}
\usepackage {graphicx}  
\usepackage{bm}        
\usepackage{amssymb}   

\begin{document}

\title{Contrasting Energy Scales of the Reentrant Integer Quantum Hall States}

\author{N. Deng$^1$, J.D. Watson$^1$, L.P. Rokhinson$^{1,2}$, M.J. Manfra$^{1,2,3}$, and G.A. Cs\'{a}thy$^1$ \footnote{gcsathy@purdue.edu}}

\affiliation{${}^1$ Department of Physics, Purdue University, West Lafayette, Indiana 47907, USA \\
${}^2$ Birck Nanotechnology Center, Purdue University, West Lafayette, Indiana 47907, USA \\
${}^3$School of Materials Engineering and School of Electrical and Computer Engineering, Purdue University, West Lafayette, Indiana 47907, USA \\}

\date{\today}

\begin{abstract}
We report drastically different
onset temperatures of the reentrant integer quantum Hall states
in the second and third Landau level. This finding is in quantitative disagreement with 
the Hartree-Fock theory of the bubble phases which is thought to describe these reentrant states.
Our results indicate that the number of electrons per bubble in either the second or the third Landau level
is likely different than predicted.
\end{abstract}

\pacs{}
\maketitle

In systems of charged particles strong Coulomb interactions stabilize a     
periodic ground state called the Wigner solid (WS) \cite{wigner,fukuyama}. The WS has been observed in
two-dimensional electron gases (2DEG) floating atop superfluids \cite{grimes},
in 2DEGs confined to GaAs/AlGaAs heterostructures \cite{2deg}, and electron bilayers \cite{bilayer}.
There is a resurgence of interest in the WS stimulated by recent work on 
electrons confined to less than two dimensions \cite{rahman}, 
2DEG in complex oxide heterostructures \cite{kozuka}, and in graphene \cite{joglekar}.
The WS was also realized in ion clouds \cite{tan} and, most recently, in cold atomic gases with dipolar interactions \cite{campbell}
and it plays a role in charged colloidal suspensions \cite{clarck} and neutron stars \cite{ogawa}.

Long range interactions may also stabilize
periodic ground states which are more intricate than the WS \cite{fogler,moessner}.
One such many-body ground state is the electronic bubble phase 
which was predicted to form in the 2DEG subjected to a perpendicular magnetic field $B$
\cite{fogler,moessner,fogler3,fogler4,haldane,yoshi,goerbig,fertig}. 
Electrons in this system move on circular Landau orbitals,
their energy is quantized to equidistant Landau levels (LL), and their ground states are labeled by the LL filling factor $\nu$ at which they form.
According to theory, the guiding centers of the Landau orbitals cluster into so called 
electron bubbles and, furthermore, the bubbles order into an isotropic lattice. Such a bubble phase can therefore be thought of 
as a WS with an internal degree of freedom, i.e. with several electrons per unit cell \cite{fogler}. 

The experimentally measured reentrant integer quantum Hall states (RIQHS) 
have been identified with the bubble phases \cite{lilly,du,eisenstein,cooper,xia,csa10,deng12,lewis,lewis2,lewis3}. 
Indeed, dc \cite{lilly,du,eisenstein,cooper,xia,csa10,deng12} and microwave transport features \cite{lewis,lewis2,lewis3}
of the RIQHSs are, generally speaking, consistent with the bubble interpretation.  
However, for the RIQHSs the number of electrons per bubble remains unknown to date. 
In lack of any direct measurements on the structure of the bubbles one has to turn to the theory. 
In the second Landau level (SLL) both two and one electron bubble phases are predicted to form 
\cite{goerbig} while in the third Landau level (TLL) only two electron bubble phases
are expected \cite{fogler3,goerbig,haldane,yoshi,fertig}. These theories, however, have their limitations.
The Hartree-Fock approach, the only one used for bubble phases both the TLL \cite{fogler3,goerbig,fertig} and the SLL \cite{goerbig}, 
is exact only in the limit of large LL occupation \cite{moessner,fogler4},
and may therefore not capture all aspects of bubbles at the lowest LL occupation, i.e. those
in the second and third LLs.
In addition, the presence of competing nearby fractional quantum Hall states in the SLL \cite{xia,csa10} is likely to
enhance fluctuations and may therefore influence electron ordering.
Finally, none of the theoretical techniques include LL mixing, an electron-electron interaction effect known
to strongly affect the energy gaps of fractional quantum Hall ground states in the SLL \cite{llm}.

Here we report sharp peaks in the temperature dependent longitudinal resistance of the RIQHSs in the TLL
which are similar to those of the RIQHSs in the SLL. This shared property highlights the common origin of these RIQHSs.
The sharp peaks allowed us to extract the onset temperatures of the RIQHSs in the TLL
which enabled a quantitative comparison of the RIQHSs forming in the TLL with those in the SLL 
as well as with the theoretically predicted bubble phases.
Our measurements of the onset temperatures are at odds with the cohesive energy calculations obtained within the 
Hartree-Fock approximation and indicate that the assignment of the RIQHSs 
to the various bubble phases is likely different than predicted.

\begin{figure*}[t]
 \includegraphics[width=2.1\columnwidth]{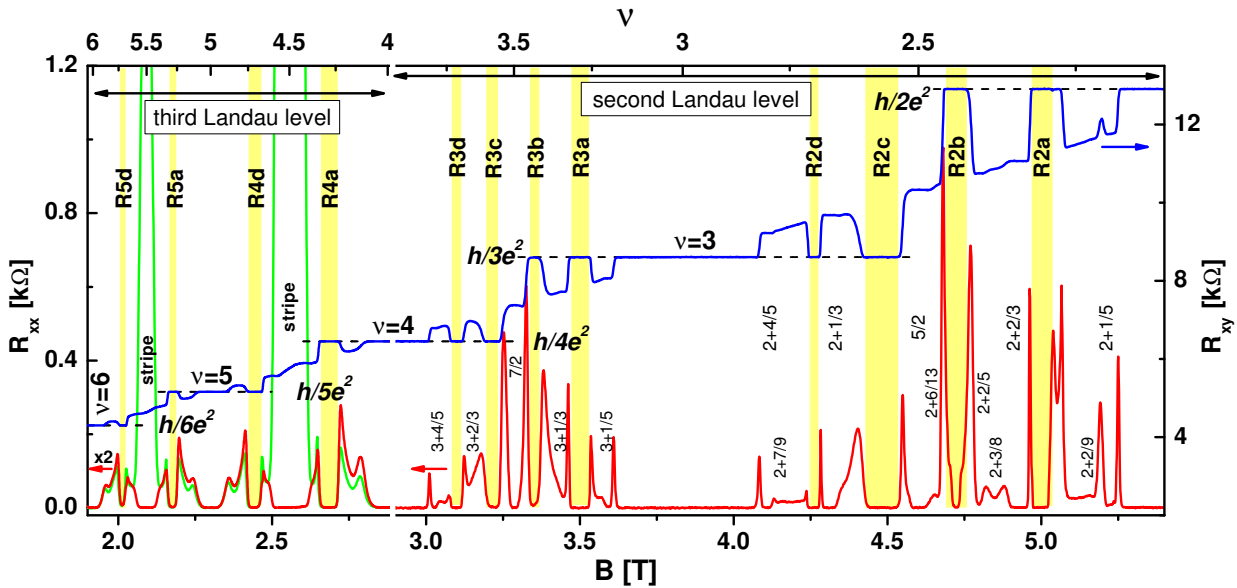}
 \caption{ The magnetoresistance in the second ($2<\nu<4$) and third ($4<\nu<6$) Landau levels
 as measured at 6.9~mK and 77mK, respectively. RIQHSs are marked by shaded stripes and FQHSs by their filling factors.
 In the TLL the two $R_{xx}$ traces shown are measured along mutually perpendicular directions and, for clarity,
 are magnified by a factor two. 
 \label{Fig1}}
 \end{figure*} 
 
We measured a high quality 2DEG confined to a 30~nm wide GaAs/AlGaAs quantum well
with a density $n=2.8\times10^{11}$cm$^{-2}$ and mobility $15\times10^{6}$cm$^{2}$/Vs grown at Purdue.
The low frequency magnetotransport measurements were performed 
at dilution refrigerator temperatures while our sample was
immersed into a liquid He-3 bath \cite{xia,csa11}. The He-3 bath facilitates cooling of the sample \cite{xia} and it enables $B$-field independent
temperature measurements by the use of a quartz tuning fork viscometer \cite{csa11}.  Due to its large heat capacity,
He-3 also serves as a thermal ballast which stabilizes the sample temperature.

In Fig.1 we show the longitudinal magnetoresistance $R_{xx}$ and the Hall resistance $R_{xy}$ plotted against $B$ and
filling factor $\nu$ in the SLL and TLL. Here $\nu=nh/eB$, where $h$ is Planck's constant and
$e$ is the elementary charge. It is important to appreciate 
that a completely filled orbital Landau level is spin-split into two distinct energy levels and, hence, its 
filling factor is $\nu=2$. Therefore the lowest Landau level corresponds to filling factors $\nu<2$, the
SLL corresponds to $2 < \nu <4$, while the TLL to $4<\nu<6$. 

The well known integer quantum Hall states are seen in Fig.1 as plateaus in $R_{xy}$ quantized to $h/ie^2$, with $i=2,3,4,5,$ and $6$.
Each of these plateaus straddle the corresponding integer filling factor $\nu=i$. As $B$ is varied, $R_{xy}$ deviates from these plateaus.
There are, however, other regions
for which $R_{xy}$ returns to an integer quantization but, in contrast to the plateaus of the integer quantum Hall states,
these plateaus develop at ranges of $\nu$ which do not contain any integer values. These features define the RIQHSs \cite{lilly,du,eisenstein,cooper}.
As an example, the RIQHS labeled $R2c$ in Fig.1 has $R_{xy}=h/3e^2$ and it stretches
between $2.54<\nu<2.60$, a region which does not contain any
integers. Quantization of $R_{xy}$ is accompanied by a vanishing $R_{xx}$.
Altogether, in the SLL there are eight RIQHS labeled $R2a$, $R2b$, $R2c$, $R2d$, $R3a$, $R3b$, $R3c$, and $R3d$ \cite{eisenstein},
while in the TLL there are only four such states labeled $R4a$, $R4d$, $R5a$, and $R5d$ \cite{lilly,du}. 
The RIQHSs are clearly marked and shaded in Fig.1.

In Fig.1 we also identify anisotropic ground states called stripe phases \cite{fogler,moessner,fogler3} 
in the vicinity of $\nu=9/2$ and $11/2$ \cite{lilly,du}, 
a very strong fractional quantum Hall state (FQHS) at $\nu=5/2$ \cite{willett} with a gap of $0.50$~K,
a well quantized $\nu=2+2/5$ FQHS,
and we discern developing FQHSs at $\nu=2+6/13$, $2+2/9$, $2+7/9$, and $2+3/8$ \cite{xia,csa10}.
We also observe a split-off RIQHS at $B$-fields exceeding that of the $R2a$ state which was
discovered in Ref.\cite{xia} and studied in detail in Ref.\cite{deng12}. 
In addition to these known aspects, we observe a new feature in the Hall resistance at $B=5.196$~T or $\nu=2.214$.
This feature is a clear deviation from the classical Hall line and it may signal the development of another RIQHS. 

\begin{table*}
 \caption{Central filling factors $\nu^*_c$ and onset temperatures $T_c$ of the RIQHSs measured. }
 \begin{ruledtabular}
 \begin{tabular}{l c c c c c c c c c c c c}
                        & $R2a$ & $R2b$ & $R2c$ & $R2d$ & $R3a$ & $R3b$ & $R3c$ & $R3d$ & $R4a$ & $R4d$ & $R5a$ & $R5d$ \\
 \hline
 $\nu^*_c$               & 0.300 & 0.438 & 0.568 & 0.700 & 0.288 & 0.430 & 0.576 & 0.713 & 0.287 & 0.714 & 0.286 & 0.714 \\
 $T_c$[mK]               & 45.3  & 29.8  & 39.9  & 29.5  & 38.1  & 25.4  & 31.0   & 25.5 & 145   & 125   & 111   & 100  \\
 \end{tabular}
 \end{ruledtabular}
\end{table*}

A notable difference between the RIQHSs is that there are twice as many of them in the SLL than in the TLL.
Despite this disparity in their numbers, the RIQHSs in the SLL and high LLs share
common features in the quantized reentrant transport \cite{lilly,du,eisenstein,cooper} and microwave response \cite{lewis,lewis2,lewis3}.
In the following we establish two additional common transport signatures of the RIQHSs in the SLL and TLL:
spikes flanking the vanishing regions of the $R_{xx}$ versus $B$ curves and 
a peak in the temperature dependent $R_{xx}$. These findings
further strengthen the argument that the RIQHSs of different LL have similar origins.

One similarity between the RIQHSs in the SLL and TLL we find is the presence of two sharp spikes in the flanks of
the vanishing region of the $R_{xx}$ versus $B$ curves, i.e. the edges of the shaded areas of Fig.1. 
Such spikes are known to be present in the flanks of the RIQHSs in the SLL \cite{xia,csa10,deng12}
and now we observe them in the TLL as well. With the exception of the data in Ref.\cite{cooper2}, 
earlier $R_{xx}$ versus $B$ curves showed a single broad peak in the region separating
the RIQHS from the nearby integer plateau; the width at half height of this peak near the RIQHSs $R4a$ 
was measured to be about 0.05~T.
In contrast, our data in Fig.1 at the corresponding fields, i.e. in the range of $2.7 \div 2.85$~T,  has a more
complex structure which exhibits a sharp spike at 2.72~T of width 0.016~T. We think that the richer structure 
in $R_{xx}$ and the presence of the sharp spikes are due to an improved sample uniformity.

\begin{figure}[b]
 \includegraphics[width=1\columnwidth]{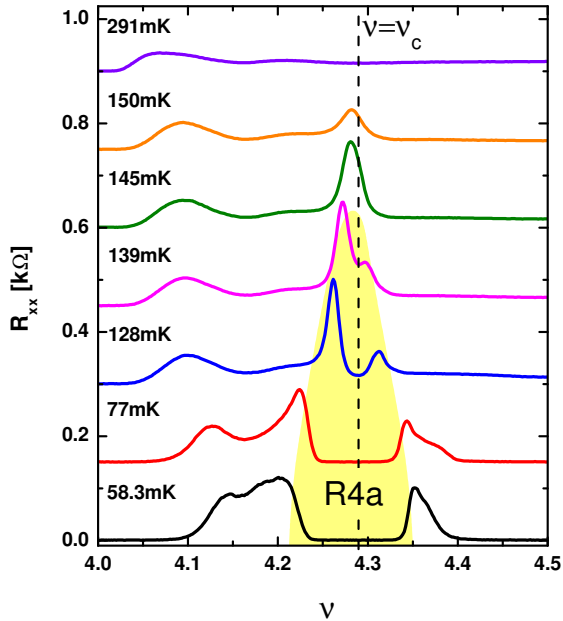}
 \caption{ The evolution with temperature of the $R4a$ RIQHS of the third Landau level. For clarity traces are shifted
 by $150\Omega$ relative to another and the reentrant region is shaded.
 \label{Fig2}}
 \end{figure}
 
Contrary to a previous report \cite{gervais}, in our sample there are no magnetoresistance 
features which may be associated with a FQHS in the TLL. We find
that the $\nu=4+1/5$ and $4+4/5$ filling factors, as seen in Fig.1 and in Fig.2, are part of the complex behavior of
$R_{xx}$ described above. Local minima do develop, but they are not located at $\nu=4+1/5$ or $4+4/5$ and, furthermore, 
they are not accompanied by a quantized Hall plateau in $R_{xy}$ (not shown) in the 
6.9 to 300~mK temperature range. Thus, in our sample there is no evidence for the formation of any FQHS in the TLL.

We find that the temperature evolution of $R_{xx}$ of the RIQHS in the TLL
and that of RIQHSs in the SLL \cite{deng12} share the following common features: 
at the lowest temperatures there are two well separated spikes of finite resistance flanking the vanishing $R_{xx}$, 
with increasing $T$ these two spike merge into a single peak, and this peak dissapears into a smooth background
with a further increase in $T$. Such a temperature dependence for the $R4a$ state of the TLL is shown in Fig.2.
We define the center of a RIQHS
as the location $\nu_c$ at which the extent of the vanishing $R_{xx}$ plateau is nearly zero. 
For example, the curve at 128~mK of Fig.2 exhibits a $R4a$ state of nearly zero width at $\nu_c=4.287$.
The partial filling factor $\nu^*_c$ is the decimal part of $\nu_c$, 
and values for the various RIQHSs are summarized in Table.I. 

\begin{figure}[b]
 \includegraphics[width=1.05\columnwidth]{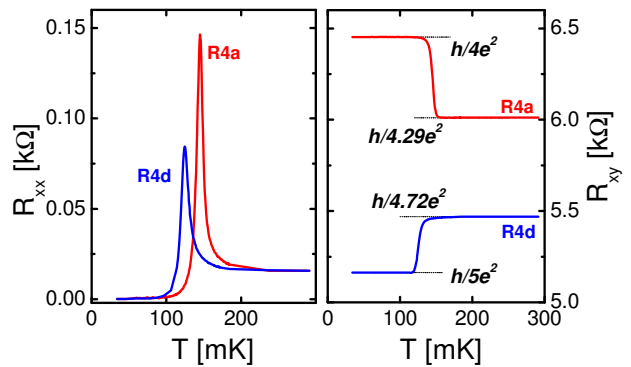}
 \caption{ The magnetoresistance $R_{xx}$ and the Hall resistance $R_{xy}$ 
 of two RIQHSs in the third Landau level measured at $\nu=4.29$ and $\nu=4.72$.
 \label{Fig3}}
 \end{figure}
 
A second shared feature of the RIQHSs in the TLL and in the SLL \cite{deng12}
is the similar $R_{xx}$ and $R_{xy}$ versus $T$ curves
measured at a fixed $\nu$. In Fig.3 we show such curves for the 
$R4a$ and $R4d$ states of the TLL in close vicinity to their respective central filling factors. 
As the temperature is increased
the Hall resistance undergoes an extremely abrupt change from the nearest integer quantized value 
to the classical Hall value $B/ne=h/\nu e^2$.
Simultaneously with the sharp change in $R_{xy}$ the longitudinal resistance
$R_{xx}$ for the $R4a$ state exhibits a sharp peak of width at half height of only $10$~mK.
We have recently reported similar dependences of both $R_{xy}$ and $R_{xx}$ of the RIQHSs in the SLL 
of a higher density sample and have interpreted
the peak temperature as the onset temperature $T_c$ of the RIQHSs \cite{deng12}. 
We thus find that a peak in the $R_{xx}$ versus $T$ curves accompanied by a sharp transition
of $R_{xy}$ from the classical Hall to a quantized value is not specific to the SLL, but 
is also a property of the RIQHSs forming in the TLL.

In the following we compare the locations, i.e. the filling factors of the RIQHSs. 
Surprisingly, the filling factors of the RIQHSs in the TLL have not yet been
measured with high precision \cite{lilly,du,eisenstein,cooper}.
Inspecting Table.I we find that $R2a$, $R3a$ from the SLL and $R4a$, and $R5a$ from the TLL
develop at similar partial filling factors. 
Indeed, $\nu^*_c|_{R3a}= \nu^*_c|_{R4a}= \nu^*_c|_{R5a}$ within our measurement error of $\pm 0.003$.
Furthermore, this common value is in close proximity to $\nu^*_c|_{R2a}$. Nonetheless, 
we measure a significant difference between  the common value of $\nu^*_c|_{Ria}$, with $i=3,4,5$ and
$\nu^*_c|_{R2a}$. This is seen in Fig.4 as an alignment
of data points associated with $R3a$, $R4a$, and $R5a$ onto a vertical dashed line
and a slight horizontal departure of the point associated with $R2a$ from this line. A similar alignment
occurs for the particle-hole symmetric states $R2d$, $R3d$, $R4d$, and $R5d$. 
We summarize thus that RIQHSs $Ria$ with $i=2,3,4$ and $5$ form at similar partial filling factors and yet
theory favors different types of order for these states: one-electron bubbles or WS at $R2a$ and $R3a$ \cite{goerbig} and 
two-electron bubbles for $R4a$ and $R5a$ \cite{fogler3,haldane,yoshi,goerbig,fertig}.

As a further test we examine the energy scales of the RIQHSs.
The cohesive energy of the bubble phase $E_{\text{coh}}$ is readily obtained from the Hartree-Fock theories \cite{fogler,moessner,fogler3,goerbig,fertig}. 
It is customary to calculate the reduced cohesive energy $e_{\text{coh}}=E_{\text{coh}}/E_c$, where
$E_c=e^2/4 \pi \epsilon l_B$ is the Coulomb energy and $l_B=\sqrt{\hbar/eB}$ the magnetic length.
Experimentally we measure the onset temperature $T_c$ and we consider the reduced onset temperature $t_c=k_B T_c/E_c$.
Fig.4 summarizes the $t_c$ of the RIQHSs in the SLL and TLL as function of $\nu^*_c$.
We assume that, within the bubble interpretation, the onset temperature of a RIQHS is a measure of its cohesive energy \cite{fukuyama}.
We find that the reduced onset temperatures $t_c$ of the RIQHSs in the SLL and TLL
are more than 2 orders of magnitude smaller than the reduced cohesive energies $e_{\text{coh}}=E_{\text{coh}}/E_c$ of the associated bubble phases 
\cite{fogler,moessner,fogler3,goerbig,fertig}.
We think this difference is most likely due to disorder and Landau level mixing
effects which are not included in the Hartee-Fock estimations \cite{fogler,moessner,fogler3,goerbig,fertig}.
Furthermore, similarly to a recent report \cite{deng12},
in the SLL we find a good collapse of $t_c$s from different spin branches and a
non-monotonic dependence of $t_c$ of $\nu_c^*$. As shown in Fig.4, $t_c$ in the TLL is in the vicinity of 
$16 \times 10^{-4}$, but the collapse of values from the two different spin branches
is not as good as for the RIQHSs in the SLL. 

Our most remarkable finding is the disproportionately large energy scale of the RIQHSs in the TLL as compared to those in the SLL.
The most striking disagreement is between the RIQHSs $R4a$ and $R2b$ believed to be two-electron bubbles.
In Ref.\cite{goerbig} the cohesive energies are calculated for both RIQHSs and they are found to be similar
$e_{\text{coh}}^{R4a} / e_{\text{coh}}^{R2b} \approx 1.2$. In contrast to these predictions,
we measure a large difference in the onset temperatures $t_{c}^{R4a} / t_{c}^{R2b} = 6.4$.
We also find $ e_{\text{coh}}^{R4a} / e_{\text{coh}}^{R2a} \approx 1$ \cite{goerbig}, while we measure $t_{c}^{R4a} / t_{c}^{R2a} = 4.3$.
In another work $e_{\text{coh}}^{R4a}$ is larger by a factor 2 \cite{fertig} as compared to that in Ref.\cite{goerbig}.
When considering $e_{\text{coh}}^{R2b}$ from Ref.\cite{goerbig},
the discrepancy between $e_{\text{coh}}^{R4a} / e_{\text{coh}}^{R2b}$ and $t_{c}^{R4a} / t_{c}^{R2b}$ is reduced by the same factor of 2, but
it still remains considerable. Taken together, we conclude that there are clear quantitative inconsistencies between the 
measured and calculated energy scales of the RIQHSs.
We note that, within the SLL, the measured and theoretical energy scales of $R2a$ and $R2b$ states compare surprisingly well:
$t_{c}^{R2a} / t_{c}^{R2b} = 1.5$ and $ e_{\text{coh}}^{R2a} / e_{\text{coh}}^{R2b} \approx 1.2$ \cite{goerbig}.

\begin{figure}[t]
 \includegraphics[width=1\columnwidth]{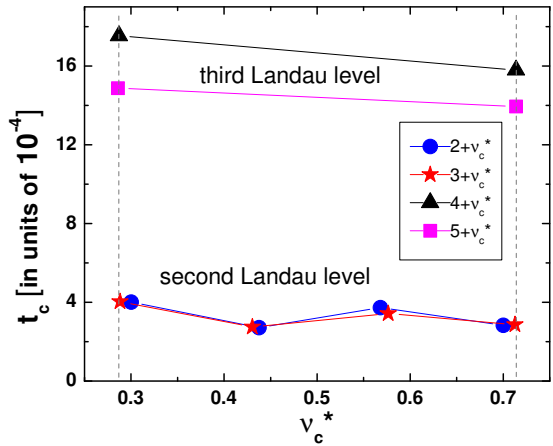}
 \caption{ The reduced onset temperatures $t_c=k_B T_c/E_c$ of the RIQHSs in the SLL and TLL  
 plotted as function of the partial filling factor $\nu^*_c$. Lines are guides to the eye.
 \label{Fig4}}
 \end{figure}

One scenario which could account for our onset temperature data is that, contrary to the
theory \cite{goerbig}, all of the RIQHSs in the SLL are bubble phases of the same type
and those in the TLL are bubbles of a different kind.
We cannot, however, discard the possibility that the RIQHSs of the second and third LLs
are the same type of bubble phases. 
The large difference in onsets could be caused by an effect dependent on LL occupancy.
Because of the presence of one extra filled LL, screening of the disorder potential in the TLL is expected to be 
more effective than that in the SLL \cite{fogler,glazman}. The substantially larger
onsets of the RIQHSs in the TLL as compared to those in the SLL could thus be a consequence of
a smoother effective disorder potential due to screening of one extra filled LL.

Finally we note that there are two recent reports of reentrant behavior in the lowest LL in 2DEGs forming in GaAs/AlGaAs hosts. 
One such observation in made in a heterostructure which has short range neutral scattering centers \cite{li}.
Another experiment was performed on wide quantum wells \cite{shay}. In both of these experiments \cite{li,shay}
reentrance has been associated with the formation of electron solids similar to the WS since
electron-electron interactions in the lowest LL are not expected to promote electronic bubble phases \cite{fogler3}.
However, the relationship of these electron solids and those in higher LLs we have studied is not understood at this time. 

To conclude, the newly reported common features in the transport of the RIQHSs both in the TLL and SLL, together with 
the reentrant behavior and radiofrequency response, supports the idea that the RIQHSs belong to the same family of ground states 
irrespective of the LL they form in. These features are qualitatively consistent with the bubble interpretation of these phases.
We found, however, that the very different energy scales of the
RIQHSs in different LLs are inconsistent with quantitative predictions of the theory of the bubbles.
This disagreement is suggestive of an assignment of the RIQHSs to bubble phases different than that proposed
by the theory. Our results call for further work in order to elucidate the nature of the RIQHSs.

This work was supported by the DOE BES grant DE-SC0006671 and we acknowledge
useful discussions with M. Fogler and Y. Lyanda-Geller.

\end{document}